\documentclass[aps,preprint,prd,showpacs,nofootinbib]{revtex4-1}
\pdfoutput=1
\usepackage{graphicx}
\usepackage{bm}
\usepackage{times}
\usepackage{hyperref}
\usepackage{slashed}
\usepackage{amsmath}
\usepackage{xcolor}
\usepackage{fancyhdr}
\usepackage{soul}
\usepackage{footmisc}

\usepackage{lipsum}
\usepackage{multirow}
\usepackage{diagbox}
\pdfoutput=1
\usepackage{amssymb}
\usepackage{hyperref}
\graphicspath{{Figures/}}

\newcommand\scalemath[2]{\scalebox{#1}{\mbox{\ensuremath{\displaystyle #2}}}}
\newcommand{\be}{\begin{equation}}
\newcommand{\ee}{\end{equation}}
\newcommand{\ba}{\begin{eqnarray}}
\newcommand{\ea}{\end{eqnarray}}
\newcommand{\bs}{\begin{subequations}}
\newcommand{\es}{\end{subequations}}

\newcommand{\grts}{\raise.3ex\hbox{$>$\kern-.75em\lower1ex\hbox{$\sim$}}}
\newcommand{\lets}{\raise.3ex\hbox{$<$\kern-.75em\lower1ex\hbox{$\sim$}}}

\newcommand{\bc}{\begin{center}}
\newcommand{\ec}{\end{center}}

\usepackage{amsmath, amssymb} 
\usepackage{slashed}
\usepackage{natbib}
\usepackage{subfigure}

\begin{document}
\title{Meson Mixing Bounds on $Z^{\prime}$ Mass in the Alignment Limit: Establishing the Phenomenological Viability of the 331 Model}
\author{Patricio Escalona$^\dagger$\footnote[1]{\href{mailto:PatricioEscalona96@gmail.com}{\footnotesize patricioescalona96@gmail.com }}}
\author{João Paulo Pinheiro$^\ddag$\footnote[2]{\href{mailto:joaopaulo.pinheiro@fqa.ub.edu}{\footnotesize joaopaulo.pinheiro@fqa.ub.edu}}}
\author{A. Doff$^\S$\footnote[3]{\href{mailto:adoff.gomes@gmail.com }{\footnotesize adoff.gomes@gmail.com }}}
\author{C. A. de S. Pires$^{\dagger}$\footnote[4]{\href{mailto:cpires@fisica.ufpb.br}{\footnotesize cpires@fisica.ufpb.br}}}

\affiliation{\footnotesize $^\dagger$Departamento de F\'isica, Universidade Federal da Para\'iba, Caixa Postal 5008, 58051-970, Jo\~ao Pessoa, PB, Brazil} 
\affiliation{\footnotesize $^{\ddag}$Departament de F\'isica Qu\`antica i Astrof\'isica and Institut de Ci\`encies del Cosmos, Universitat de Barcelona, Diagonal 647, E-08028 Barcelona, Spain}
\affiliation{\footnotesize $^{\S}$ Universidade Tecnologica Federal do Parana - UTFPR - DAFIS, R. Doutor Washington Subtil Chueire, 330 - Jardim Carvalho, 84017-220, Ponta Grossa, PR, Brazil}

\vspace{1cm}
\begin{abstract}
We perform a systematic study of flavor-changing neutral currents (FCNCs) in the 331 model with right-handed neutrinos (331RHNs), analyzing constraints on the $Z^\prime$ boson mass from $K$-, $D$-, $B_d$-, and $B_s$-meson oscillations. By explicitly incorporating scalar sector dynamics and quark rotation ambiguities ($V_L^{u,d}$), we demonstrate that $Z'$ mass limits depend critically on the parametrization of Cabibbo-Kobayashi-Maskawa (CKM) matrix factors. Three scenarios are explored: \textbf{(i)} $V_L^u = V_\text{CKM}^\dagger$ (FCNCs restricted to $D$-mesons), \textbf{(ii)} $V_L^d = V_\text{CKM}$ (dominant $B_s$ constraints), and \textbf{(iii)} a hybrid mixing pattern. Strikingly, \textbf{scenario i)} reduces the $Z'$ mass bound to $M_{Z'} \gtrsim 600\;\text{GeV}$ at leading order and $M_{Z'} \gtrsim 473\;\text{GeV}$ at next-to leading order—two orders of magnitude below literature values—by leveraging large experimental uncertainties in $D$-$\bar{D}$ oscillations. Conversely, \textbf{scenario ii)} requires $M_{Z'} \gtrsim 165\;\text{TeV}$ due to stringent $B_s$ data. We further establish the alignment limit $\cos(\phi+\varphi) = 0$ for the SM-like Higgs, showing its viability depends on $V_L^{u,d}$ configurations, with $B_s$ systems enforcing $|\cos(\phi+\varphi)| < 0.01$ in down-sector FCNC scenarios. Our analysis reveals that strategic choices of quark mixing matrices can suppress FCNC visibility, reconciling the 331 framework with flavor data without ultra-heavy $Z'$ bosons. This work provides the first unified treatment of SM-like Higgs- and $Z'$-mediated FCNCs in 331 models, identifying viable parameter spaces for collider phenomenology.
\end{abstract}
\maketitle

\section{Introduction}
The Glashow-Iliopoulos-Maiani (GIM) mechanism \cite{Glashow:1970gm} suppresses Flavor-Changing Neutral Current (FCNC) processes in the Standard Model (SM). Due to this mechanism, these processes are absent at tree-level and highly suppressed at one-loop. This suppression arises from the unitarity of the Cabibbo-Kobayashi-Maskawa (CKM) matrix, making FCNC-sensitive observables like meson-antimeson oscillations ($K^0$-$\bar{K}^0$, $B^0_s$-$\bar{B}^0_s$, etc.) exceptionally precise probes of physics beyond the SM (BSM). Experimental measurements of these oscillations, particularly in $B_s$- and $K$-meson systems \cite{LHCb:2021moh}, align with SM predictions at the $\mathcal{O}(1\%)$ level, leaving minimal room for new physics (NP) contributions. Consequently, BSM models must either replicate SM loop-level predictions or introduce mechanisms to evade these stringent constraints (see, e.g. Ref.~\cite{Buras:2023fhi}). 

Among NP frameworks, 331 models \cite{Singer:1980sw,Frampton:1992wt,Pisano:1992bxx,Foot:1994ym,Long:2024gyy}—gauge extensions of the SM based on the $SU(3)_C \times SU(3)_L \times U(1)_X$ symmetry—naturally incorporate tree-level FCNCs \cite{Montero:1992jk}. Anomaly cancellation in these models requires non-universal quark family assignments: one generation transforms as an $SU(3)_L$ triplet, while others are antitriplets. This asymmetry misaligns quark mass and interaction eigenstates, generating tree-level FCNCs mediated by a heavy $Z^{\prime}$ boson \cite{Frampton:1992wt,Ng:1992st,Liu:1993gy,GomezDumm:1993oxo,Pisano:1996ht}. Current limits on meson oscillations constrain the $Z^{\prime}$ mass to multi-TeV scales (e.g., $M_{Z^{\prime}} \gtrsim 1$–$100$ TeV) \cite{Long:1999ij,Martinez:2008jj,Buras:2012dp,Cogollo:2012ek,Queiroz:2016gif,Okada:2016whh,CarcamoHernandez:2022fvl, deJesus:2023lvn,Oliveira:2022dav,Oliveira:2022dav,Buras:2023ldz,Buras:2023fhi,Huitu:2024nap}, highlighting the tension between 331 models and flavor data. However, existing studies predominantly focus on $Z^{\prime}$-mediated processes, neglecting the scalar sector’s role—a critical lapse given its inherent capacity to induce FCNCs. 

The scalar sector of 331 models, responsible for breaking $SU(3)_L \times U(1)_X$ to the SM electroweak symmetry, includes multiple Higgs fields \cite{Foot:1994ym}. These scalars not only generate fermion masses but also introduce flavor-violating Yukawa couplings due to the family-dependent $SU(3)_L$ assignments. Crucially, the observed CKM matrix emerges as a product of the left-handed quark rotation matrices $V_{\text{CKM}} = V_L^u (V_L^d)^\dagger$ where $V_L^{u,d}$ are unitary matrices that diagonalize the up- and down-type quark mass matrices, respectively. Unlike $Z^{\prime}$-mediated FCNCs, scalar-induced flavor transitions cannot be suppressed by simply increasing the mediator mass \cite{Cogollo:2012ek,Okada:2016whh,Oliveira:2022vjo}: the scalar couplings are tied to quark masses and CKM parameters, while at least one of them should play the role of the SM Higgs, required to mimic the observed $125$ GeV boson and inevitably acquires flavor-violating couplings, mediating processes like $D^0$-$\bar{D}^0$ mixing via $c \leftrightarrow u$ transitions. Such contributions scale as $1/m_h^2$, conflicting with experimental bounds unless finely tuned. Conversely, as we will discuss in this paper, the extended scalar sector accommodates the \textit{alignment limit}, albeit under conditions \textbf{distinct} from those in the conventional Two Higgs Doublet Model (2HDM) \cite{Branco:2011iw, Gunion:2002zf}. For the first time in the literature, a detailed exploration of these modified conditions, along with their implications, will be presented in this work.  Importantly, the structure of $V_L^{u,d}$—and their impact on $V_{\text{CKM}}$—plays a crucial role in determining the magnitude of FCNCs. By parameterizing these rotations, one can systematically explore how deviations from specific $V_L^{u,d}$ configurations alter predictions for meson oscillations and, consequently, the inferred lower bounds on the $Z^{\prime}$ mass.

This work addresses this gap by analyzing the scalar sector’s impact on FCNC observables and its interplay with $Z^{\prime}$-mediated processes. We demonstrate that the interplay between $V_L^{u,d}$ rotations and scalar/gauge couplings imposes bounds on the $Z'$ mass that depend critically on the parameterization of the quark mixing matrices. For instance, certain $V_L^{u,d}$ configurations suppress $Z'$-mediated contributions to $B_s$-$\bar{B}_s$ mixing, relaxing $M_{Z^{\prime}}$ constraints, while others exacerbate tensions with data. By scanning over viable $V_L^{u,d}$ parametrizations, we determine the allowed upper and lower bounds on $Z^{\prime}$ mass limits via meson oscillation observables.  

By scanning over viable $V_L^{u,d}$ parametrizations, we identify the most stringent allowed bounds on the $Z^{\prime}$ mass, reducing previous limits to $\sim 600\;\text{GeV}$ at leading order (LO) (and $\sim 473\;\text{GeV}$ at next-to leading order (NLO)) through optimized choices of the quark rotation matrices $V_L^{u,d}$. Notably, we establish the alignment limit in the scalar sector of the $\text{331}$ model and demonstrate its explicit dependence on the $V_L^{u,d}$ parametrization. While our study focuses on the variant where the third fermion family transforms distinctively under the gauge group, we emphasize that alternative $\text{331}$ model variants—coupled with distinct $V_L^{u,d}$ rotations—would yield different bounds on both the alignment limit and the $Z'$ mass. These results underscore the necessity of a comprehensive framework for $\text{331}$ models, as scalar-sector dynamics and quark rotation ambiguities critically govern their viability against precision flavor data. A systematic exploration of these correlations, including their phenomenological implications, will be detailed in subsequent sections.

In Section \ref{sec:model}, we introduce the model and the main considerations about the rotation matrices. In Section \ref{sec:FCNC_higgs} we introduce the FCNC interactions induced by the SM-like Higgs, present the alignment condition, and the main results after comparing with recent bounds over meson oscillation parameters. In Section \ref{sec:FCNC_Zp},  we introduced the FCNC interactions induced by the SM-like Higgs and the bounds over the mass of the $Z^{\prime}$ for each choice of parametrization of the CKM matrix, before we conclude in Section \ref{sec:summary}. The appendix contains details on the analysis like the neutral scalar sector of the 331 in Appendix~\ref{app:scalarpotential} and the coefficients that contribute to FCNC processes of the scalar section in Appendix~\ref{app:coefficients}.

\section{The model}
\label{sec:model}
Our investigation of FCNC processes will be placed in a particular version of the  $SU(3)_C \times SU(3)_L \times U(1)_N$ gauge symmetry model called 331 with right-handed neutrinos \cite{Foot:1994ym}. Here, right-handed neutrinos compose the triplet of leptons:
\begin{equation}
f_{aL}= \begin{pmatrix}
\nu_{a}     \\
\ell_{a}       \\
\nu^{c}_{a} \\
\end{pmatrix}_{L} \sim (1,3,-1/3), \quad e_{aR}\sim (1,1,-1),
\end{equation}
with $a=e\,,\,\mu\,\,\tau$ representing the three SM generations of leptons.

Regarding the hadronic sector, gauge anomaly cancellation requires that the quark generations transform differently under $SU(3)_L$ group \cite{Frampton:1992wt,Ng:1992st,Liu:1993gy,Pisano:1996ht}. Usually, the third family is adopted to be the one that transforms differently (due to the heaviness of the top-quark mass ) \cite{Ng:1992st,Long:1999ij,Oliveira:2022vjo}. This choice is called variant I \cite{Oliveira:2022vjo}, that will be adopted in our studies (a detailed study of the variants of the 331 is presented in Ref.~\cite{Oliveira:2022vjo}). In this variant, the third generation transforms as triplets under $SU(3)_L$,
whereas the first and second ones as anti-triplets as follows,
\begin{eqnarray}
&&Q_{i_L} = \left (
\begin{array}{c}
d_{i} \\
-u_{i} \\
d^{\prime}_{i}
\end{array}
\right )_L\sim(3\,,\,\bar{3}\,,\,0)\,,u_{iR}\,\sim(3,1,2/3),\,\,\,\nonumber \\
&&\,\,d_{iR}\,\sim(3,1,-1/3)\,,\,\,\,\, d^{\prime}_{iR}\,\sim(3,1,-1/3),\nonumber \\
&&Q_{3L} = \left (
\begin{array}{c}
u_{3} \\
d_{3} \\
u^{\prime}_{3}
\end{array}
\right )_L\sim(3\,,\,3\,,\,1/3),u_{3R}\,\sim(3,1,2/3),\nonumber \\
&&\,\,d_{3R}\,\sim(3,1,-1/3)\,,\,u^{\prime}_{3R}\,\sim(3,1,2/3),
\label{quarks} 
\end{eqnarray}
where  the index $i=1,2$ is restricted to the first two generations.  The negative signal in the anti-triplet $Q_{i_L}$ is just to standardize the signals of the charged current interactions with the gauge bosons.  The primed quarks are new heavy quarks with the usual $(+\frac{2}{3}, -\frac{1}{3})$ electric charges. From Eq.~\eqref{quarks}, it is clear that 331 models naturally account for FCNC processes, since it treats one family differently from the others.

Fermion masses are generated through the spontaneous symmetry-breaking (SSB) mechanism governed by three scalar triplets, $\eta$, $\rho$ and $\chi$ \cite{Foot:1994ym},
\begin{eqnarray}
&&\eta = \left (
\begin{array}{c}
\eta^0 \\
\eta^- \\
\eta^{\prime 0}
\end{array}
\right ),\,\rho = \left (
\begin{array}{c}
\rho^+ \\
\rho^0 \\
\rho^{\prime +}
\end{array}
\right ),\,
\chi = \left (
\begin{array}{c}
\chi^0 \\
\chi^{-} \\
\chi^{\prime 0}
\end{array}
\right ),
\label{fermiontriplets} 
\end{eqnarray}
with $\eta$ and $\chi$ transforming as $(1\,,\,3\,,\,-1/3)$, $\rho$ as $(1\,,\,3\,,\,2/3)$. Three neutral components of the scalar triplets develop vev, as:
\begin{equation}
  \eta^0\,\,, \rho^0\,\,, \chi^{\prime 0} =\frac{1}{\sqrt{2}}(v_{\eta\,, \rho\,,\chi^{\prime}}+R_{_{\eta\,, \rho\,,\chi^{\prime}}}+iI_{_{\eta\,, \rho\,,\chi^{\prime}}})\,,
\label{eq:expansion_neutralscalars}
\end{equation}
The details of the potential are described in Appendix~\ref{app:scalarpotential}. The fermion masses will be generated after the SSB of the neutral scalars,  $\langle \eta^0 \rangle =v_\eta\,\,, \langle \rho^0 \rangle=v_\rho$ and $\langle \chi^{\prime 0} \rangle=v_{\chi'}$, yielding masses for the SM quarks 
through the Yukawa interactions:
\begin{eqnarray}
&-&{\cal L}^Y \supset   g_{ia}\bar Q_{i_L}\eta^* d_{a_R} +h_{3a} \bar Q_{3_L}\eta u_{a_R} +g_{3a}\bar Q_{3_L}\rho d_{a_R}+h_{ia}\bar Q_{i_L}\rho^* u_{a_R}+   \mbox{H.c} \,,
\label{yukawa}
\end{eqnarray}
where $a=1,2,3$. \footnote{ Notice that this Yukawa interaction is the simplest version of the model. It is achieved by requiring the Lagrangian to be symmetric under a $Z_2$ discrete symmetry, with the fields transforming as
\begin{equation}
(\rho\,,\,\chi\,,\,e_{a_R}, u_{a_R}, Q_{3_L}\,,\, d^{\prime}_{i_L}) \to -(\rho\,,\,\chi\,,\,e_{a_R}, u_{a_R}, Q_{3_L}\,,\, d^{\prime}_{i_L}).
\label{Z2}
\end{equation}
} It is important to note that in this variant the mass of the top quark is dominantly determined by the $v_\eta$. 

Now, assuming that the right-handed quarks come in a diagonal basis, it is possible to associate the Yukawa couplings $g$ and $h$, with the standard quarks masses and mixing by the following identity \cite{Oliveira:2022vjo}:
\begin{eqnarray}
\label{eq:yukawa_couplings}
   && g_{ia}=\sqrt{2}(V^d_L)_{ia}\frac{(m_{\text{down}})_a}{v_\rho}\,, \, g_{3a}=\sqrt{2}(V^d_L)_{3a}\frac{(m_{\text{down}})_a}{v_\rho}\,,\nonumber \\
   &&  h_{ia}=-\sqrt{2}(V^u_L)_{ia}\frac{(m_{\text{up}})_a}{v_\eta}\,, \, h_{3a}=\sqrt{2}(V^u_L)_{3a}\frac{(m_{\text{up}})_a}{v_\eta},
\end{eqnarray}
where $i=1,2$ and $a=1,2,3$ with  $(m_{\text{down}})_{(1,2,3)}=(m_d, m_s,m_b)$ and  $(m_{\text{up}})_{(1,2,3)}=(m_u, m_c,m_t)$. The mixing matrices  $V^{d,u}_L$  bring the  quarks in the symmetric basis, $u=(u_1 \,,\, u_2 \,,\, u_3)^T_L$ and $d=(d_1 \,,\, d_2 \,,\, d_3)^T_L$,  to the physical ones, $\hat u=(u\,,\, c \,,\, t)^T_L$ and $\hat d=(d\,,\, s\,,\, b)^T_L$ .  

In this paper, we adopt the assumption that right-handed quark mixing is absent. This is formalized by setting the right-handed quark mixing matrices to the identity matrix:
\begin{equation}
\label{eq:quarks_rh}
V_R^d = V_R^u = \mathbf{1},
\end{equation}
where $\mathbf{1}$ denotes the identity matrix in flavor space. This simplification implies that the observed quark mixing arises solely from the left-handed sector.
The left-handed quark mixing is governed by two unitary matrices, $V^{u,d}_L$, which rotate the quark fields from their flavor eigenstates to their mass eigenstates. These matrices satisfy the following relations:
\begin{equation}
\label{eq:par_CKM}
V_L^{u \dagger} V_L^d = V_\text{CKM}, \quad
V_L^{u \dagger} V_L^u = \mathbf{1}, \quad
V_L^{d \dagger} V_L^d = \mathbf{1},
\end{equation}
where $V_\mathrm{CKM}$ is the CKM matrix. The first equality ensures that the product of the left-handed mixing matrices reproduces the experimentally observed CKM matrix, which parametrizes weak interaction flavor mixing in the SM. The most recent bounds over the CKM matrix components are \cite{PDG2024}:
\begin{equation}|V_{\text{CKM}}|=
\begin{pmatrix}
0.97435 \pm 0.00016 & 0.22501 \pm 0.00068 & 0.003732^{+0.000090}_{-0.000085} \\
0.22487 \pm 0.00068 &  0.97349 \pm 0.00016 &  0.04183^{+0.00079}_{-0.00069} \\
0.00858^{+0.00019}_{-0.00017} & 0.04111^{+0.00077}_{-0.00068} & 0.999118^{+0.000029}_{-0.000034}
\end{pmatrix}.
\label{eq:CKM_values}
\end{equation}
The latter two equalities in Eq.~\ref{eq:par_CKM} reflect the unitarity of $V_L^{d,u}$. 

Now, that we have described the model, we are able to search for the phenomenological consequences of the FCNC interactions mediated by the SM-like Higgs and its main consequences in meson-antimeson oscillation.

\section{FCNC induced by the SM-like Higgs}
\label{sec:FCNC_higgs} 
As described in the last section, the Yukawa interaction in Eq.~\eqref{yukawa} leads to FCNC interactions mediated by the neutral scalars. To properly analyze these effects, the scalar fields must first be rotated to their mass eigenstates. A detailed discussion of the scalar sector mixing mechanism is provided in Appendix~\ref{app:scalarpotential}. In short, the scalar spectrum consists of one massive pseudoscalar $A$ and three CP-even scalars: $H'$, $H$, and $h$. While $H'$ predominantly originates from the $R_{\chi'}$ component and decouples from the other scalars, the $A$, $H$, and $h$ particles form a scalar spectrum characteristic of many SM extensions, such as 2HDM. 

In this framework, the $A$ and $H$ scalars are heavy\footnote{These particles cannot be lighter than $350$ GeV \cite{Cherchiglia:2022zfy}.} and mass-degenerate with 
\begin{equation}
m_H = m_A = m = 10\ \text{TeV},
\end{equation}
while $h$ corresponds to the SM-like Higgs boson with a mass of 
\begin{equation}
m_h = 125.11\ \text{GeV}.
\end{equation} Such choice is to avoid any relevant contribution to FCNC observables from the new set of scalars. 

Having established the scalar mass hierarchy, we now analyze FCNC processes mediated by Higgs bosons. 
The SM-like Higgs $h$ emerges as a mixture of the CP-even components from the $\eta^0$ and $\rho^0$ scalar fields, parametrized by
\begin{equation}
    h = \cos\varphi\, R_\eta + \sin\varphi\, R_\rho,
    \label{eq:higgs_mixing}
\end{equation}
where $\varphi$ represents the scalar mixing angle. This decomposition enables us to derive the Yukawa interaction Lagrangian between the Higgs field $h$ and the physical quark states that contributes for FCNC processes at tree-level:
\begin{eqnarray}
{\cal L}_Y^{h}&=&  \frac{2 }{|\sin(2\phi)| }    \left[  \cos(\phi+\varphi) (V^u_L)_{3a}^* (V^u_L)_{b3} \right]\frac{(m_{\text{up}})_a}{v} \bar{\hat{u}}_{b_L} \hat u_{a_R}h  \nonumber \\
&-&\frac{2 }{|\sin(2\phi)| }   \left[ \cos(\phi+\varphi)   (V^d_L)_{3a}^* (V^d_L)_{b3} \right]\frac{( m_{\text{down}})_a}{v} \bar{\hat{d}}_{b_L} \hat d_{a_R}h + H.c.,
\label{eq:Ycaseh1}
\end{eqnarray}
where the indexes $i=1,2$ and $a,b=1,2,3$
\footnote{The remaining Yukawa interactions with the scalars $A$ and $H$ are present in Appendix~\ref{app:coefficients}.}.
As notation we use   $s_\varphi=\sin{\varphi}$ and $c_\varphi=\cos{\varphi}$. We define  $v^2_\eta + v^2_\rho =v^2$, with $v$ being the standard vev, ($v=246$ GeV),  and   $\tan \phi=\frac{v_\eta}{v_\rho}$.
\footnote{In this paper, we are assuming a different notation of the typical 2HDM. There, $\tan \beta$ is the ratio between the vevs and $\alpha$ is similar to what we call $\varphi$. The reason of the change is straightforward. The alignment limit of the 2HDM is known as $\cos(\beta-\alpha)\to 0$, while the alignment of the 331 is slightly different, $\cos(\phi+\varphi)\to 0$. Additionally, $\tan\beta$ of the 2HDM is equivalent to our $\cot\phi$. Then, in order to avoid misinterpretation of our results, we are adopting a different nomenclature for the mixing angles.}

Three crucial features emerge from the Lagrangian in Eq.~\eqref{eq:Ycaseh1}:

\begin{itemize}    
\item \textbf{I:} The FCNC interactions arises exclusively through combinations of the mixing matrix elements $(V_L^{u,d})_{3a}^*$ and $(V_L^{u,d})_{b3}$. This pattern originates from our choice of \textbf{Variant 1}, where the third quark family transforms under a different gauge group representation. The index 3 reflects this family-specific transformation property. For alternative scenarios (Variants 2 or 3), the relevant indices would shift accordingly while maintaining analogous matrix element dependencies.    
    \item \textbf{II:} The FCNC interaction mediated by $h$ depends on the amplification factor $f(\phi) \equiv 1/|\sin(2\phi)|$. This function reaches its minimum value $f(\pi/4) = 1$ when $\tan\phi = 1$, corresponding to maximal mixing ($\phi = \pi/4$). Any deviation from $\tan\phi = 1$ (either $\tan\phi > 1$ or $\tan\phi < 1$) enhances $f(\phi)$, thereby amplifying FCNC contributions. This establishes $\tan\phi =\frac{v_\eta}{v_\rho} =1$ as the \textbf{optimal parameter choice for FCNC suppression}.
    \item \textbf{III:} The critical dependence on $\cos(\phi + \varphi)$ reveals a novel alignment mechanism. The condition $\cos(\phi + \varphi) = 0$ completely decouples the SM-like Higgs $h$ from FCNC processes, defining the \textit{alignment limit} of the 331 model. Our work presents the first systematic derivation of this alignment condition within the 331 framework, establishing
    \begin{equation}
        \phi + \varphi = \frac{\pi}{2}
    \end{equation}
    as the crucial relationship for Higgs-mediated FCNC suppression.
\end{itemize}

A direct consequence of FCNC interactions is to measure the contribution at tree-level to meson oscillation observables. The Higgs-like particle contribution to the mass difference is
\begin{align}
\Delta  M_{M}^{h}=8\Bigg(\frac{ \cos(\phi+\varphi)}{\sin  2\phi}\Bigg)^2\frac{m_{M} B_{M} f_{M}^2}{m_{h}^2}\left[\frac{5}{24} \operatorname{Re}\left[\left(C_{M,h}^L\right)^2+\left(C_{M,h}^R\right)^2\right]\left(\frac{m_{M}}{m_{q_a}+m_{q_b}}\right)^2 \right. \nonumber \\  + \left( 2 \operatorname{Re}\left[C_{M,h}^L C_{M,h}^R\right] \left(\frac{1}{24} +\frac{1}{4}\left(\frac{m_{M}}{m_{q_a}+m_{q_b}}\right)^2\right)\right]\label{contribution_h},
\end{align}
where the $M$ stands for the oscillated meson, and $m_M,B_M$ and $f_M$ are the mass, bag parameter, and decay constant of the respective meson. The coefficients $C_{M,h}^{L,R}$ are explicitly written in the Appendix \ref{app:coefficients}. The calculation follows the procedure of Ref.~\cite{Gabbiani:1996hi}.

\begin{table}[ht!]
\centering
\begin{tabular}{|c|c|c|c|}
\hline
Meson (M)  & $\Delta M_M^{\text{exp}} \pm \delta\Delta M_M^{\text{exp}}$ (MeV) & $M_M$ (MeV) & $\sqrt{B_M}f_M$ (MeV) \\
\hline
$D$ &  $(6.562 \pm 0.764) \times 10^{-12}$ \cite{PDG2024} & $1864.84 \pm 0.05$ \cite{PDG2024} & $212.7 \pm 8.3$ \cite{FLAG} \\
\hline
$K$ &  $(3.484 \pm 0.006) \times 10^{-12}$ \cite{PDG2024} & $497.611 \pm 0.013$ \cite{PDG2024} & $156.3 \pm 0.6$ \cite{FLAG} \\
\hline
$B_d$ &  $(3.336 \pm 0.013) \times 10^{-10}$ \cite{PDG2024} & $5279.72 \pm 0.08$ \cite{PDG2024} & $225 \pm 9$ \cite{FLAG} \\
\hline
$B_s$ &  $(1.1693 \pm 0.0004) \times 10^{-8}$ \cite{PDG2024} & $5366.93 \pm 0.10$ \cite{PDG2024} & $274 \pm 8$ \cite{FLAG} \\
\hline
\end{tabular}
\caption{Experimental parameters for neutral meson mixing.}
\label{tab:data_table}
\end{table}

We are constraining the effects of the FCNC interaction mediated by $h$ using current experimental results for meson oscillations. As discussed in the introduction and shown in Table \ref{tab:data_table}, SM predictions for $\Delta m_{M}$ agree with measurements less than $2\sigma$ range. Consequently, new physics contributions must not exceed experimental errors at this uncertainty: $\Delta M_{M}^{331}\leq \delta\Delta M_{M}^{\text{exp}}$.

Note that the contribution of $h$ is not only modulated by the mixing angles of the scalar sector, $\varphi$ and $\phi$, but also crucially depends on the patterns of quark mixing, that is, on the entries of the matrices $V_L^u$ and $V_L^d$. 
Any phenomenologically consistent FCNC study in this context must respect the $V_\text{CKM}$ factorization in $V_L^u$ and $V_L^d$. This relation imposes non-trivial correlations between the 4 relevant mesons.

On the other hand, note that neither this requirement nor unitarity of individual matrices is sufficient to unambiguously define $V_L^u$ and $V_L^d$. However, among the infinite possible quark mixing scenarios, two trivial cases stand out: either $V_L^u$ or $V_L^d$ can be set equal to the identity matrix, leaving the other as the CKM matrix. These cases represent two extreme yet illustrative scenarios of quark mixing patterns. In what follows, the contributions $\Delta M_{M}^{h}$ and its dependence on scalar mixing configuration are described in detail for these scenarios.
 
\textbf{Scenario (i):} Consider the scenario in which $V_L^u = V_{\text{CKM}}^\dagger$. For this choice of quark mixing, the contribution of $h$ to the mass differences of $K$, $B_s$, and $B_d$ mesons is exactly zero, so the mass difference $\Delta M_D^h$ becomes the only relevant quantity. On the other hand, the relative uncertainty of the measurement of the $D-\bar{D}$ 
 oscillation is the largest among mesons (see Table~\ref{tab:data_table}). This configuration consequently produces the \textit{weakest possible constraints} over the scalar mixing parameters $\phi$ and $\varphi$, establishing it as the most phenomenologically flexible scenario for 331 model building (as we will see in Section \ref{sec:FCNC_Zp}, this scenario gives the most unconstrained bound over the mass of the boson $Z'$ ). 
 
\begin{figure}[h!]
    \centering    \includegraphics[width=0.49\linewidth]{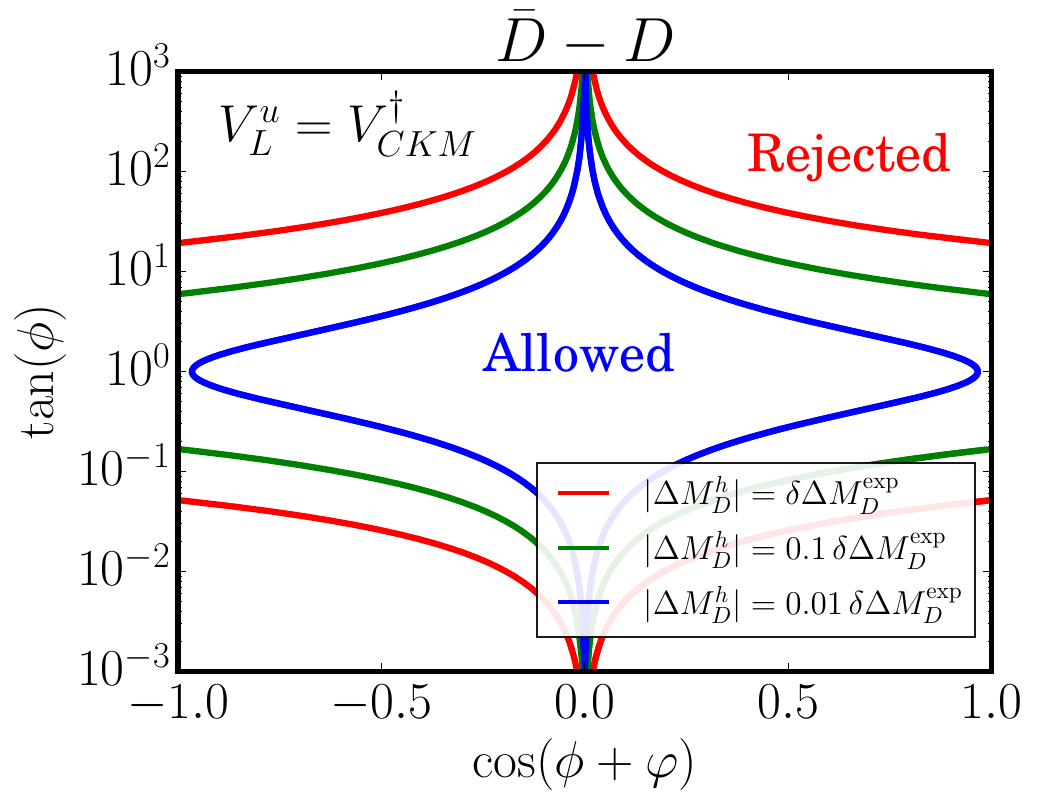}
\includegraphics[width=0.49\linewidth]{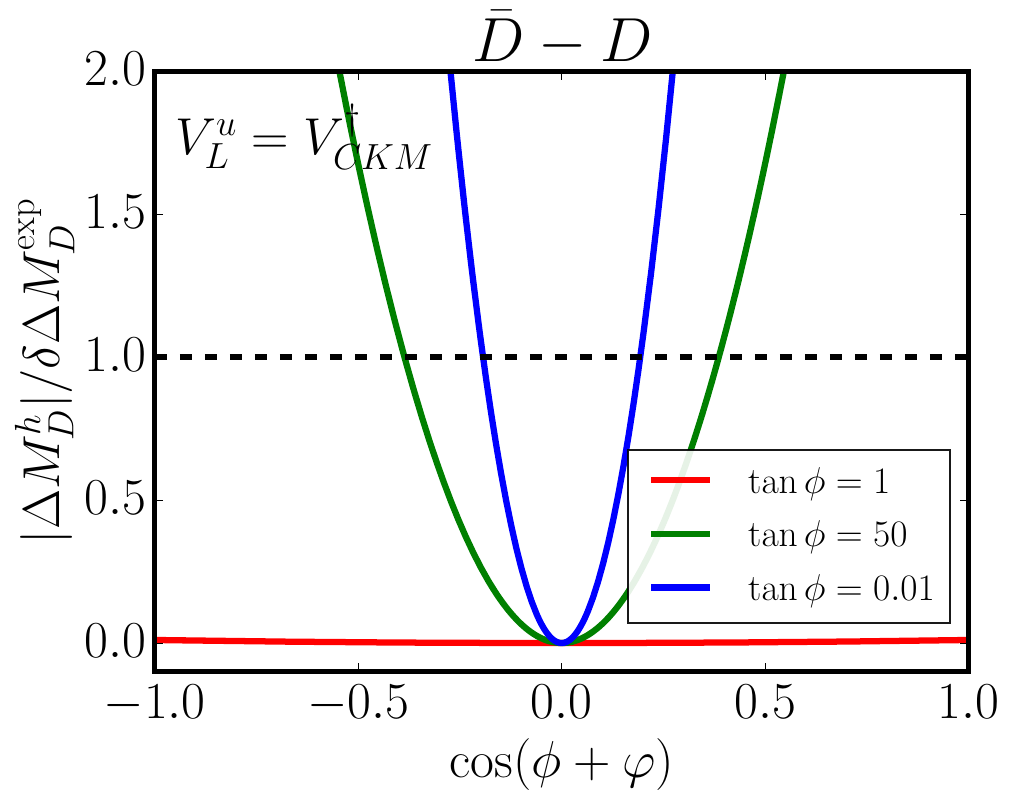}
    \caption{(a) Contours on the $\cos(\phi+\varphi)$--$\tan\phi$ plane fulfilling $100\%$ (red), $10\%$ (green) and $1\%$ (blue) of the mass difference $\delta\Delta M_D ^\text{exp}$. As indicated in the figure, outwards of the contours represents excluded region of the parameter space. Here, the mixing pattern is $V_L^u=V_\text{CKM}^\dagger$ and $V_L^d=\mathbf{1}$. (b) Mass difference as a function of $\cos(\phi+\varphi)$ for $\tan(\phi)=1$ (red), $\tan\phi=50$ (green) and $\tan\phi = 0.01$ (blue). The dashed black line denotes $|\Delta M_D^h|=\delta \Delta M_D^\text{exp}$.}
    \label{fig:higgs_i}
\end{figure}

In Fig.~(\ref{fig:higgs_i}a), we present the parameter space dependence of the Higgs-mediated contribution $\Delta M^h_M$ in the $\cos(\phi+\varphi)$--$\tan\phi$ plane. The exclusion regions are demarcated by three contours: red (for $\Delta M_D^h/\delta\Delta M_D^\text{exp} = 1$), green (for $\Delta M_D^h/\delta\Delta M_D^\text{exp} = 0.1$), and blue (for $\Delta M_D^h/\delta\Delta M_D^\text{exp} = 0.01$). Regions interior to the curves are experimentally allowed, while exterior regions are excluded. Notably, for $\Delta M_D^h/\delta\Delta M_D^\text{exp} = 1$ (red curve), no alignment limit constraints arise for $\tan\phi > 20$ or $\tan\phi < 0.05$. Tightened bounds emerge when restricting the Higgs contribution to $10\%$ or $1\%$ of the total experimental uncertainty (green and blue curves, respectively).

In Fig.~(\ref{fig:higgs_i}b), we analyze the $\tan\phi$ dependence across the $\cos(\phi+\varphi)$--$\Delta M^h_M$ parameter space. Three benchmark scenarios are shown: $\tan\phi = 1$ (red), $\tan\phi = 50$ (green), and $\tan\phi = 0.01$ (blue). The distinct trajectories of these curves illustrate how variations in $\tan\phi$ reshape the allowed regions, particularly at extreme values of $\cos(\phi+\varphi)$. This behavior reviews the interplay between quark rotation parametrizations ($V_L^{u,d}$) and scalar alignment conditions in determining viable $\text{331}$ model parameter spaces. 

\textbf{Scenario (ii)}: Consider the scenario in which $V_L^d = V_{\text{CKM}}$. For this choice of quark mixing, the contribution of $h$ to the mass differences of D mesons is exactly zero; then there are FCNC contributions to $K$, $B_s$, and $B_d$ meson oscillations. In opposition to \textbf{scenario i)}, the relative uncertainty of the measurement of the $B_s-\bar{B_s}$ 
 oscillation is the smallest among mesons (see Table~\ref{tab:data_table}). This configuration consequently produces another possible constraint over the scalar mixing parameters $\phi$ and $\varphi$, establishing it as the most phenomenologically flexible scenario for 331 model building.  
 
\begin{figure}[ht!]
    \centering
\includegraphics[width=0.49\linewidth]{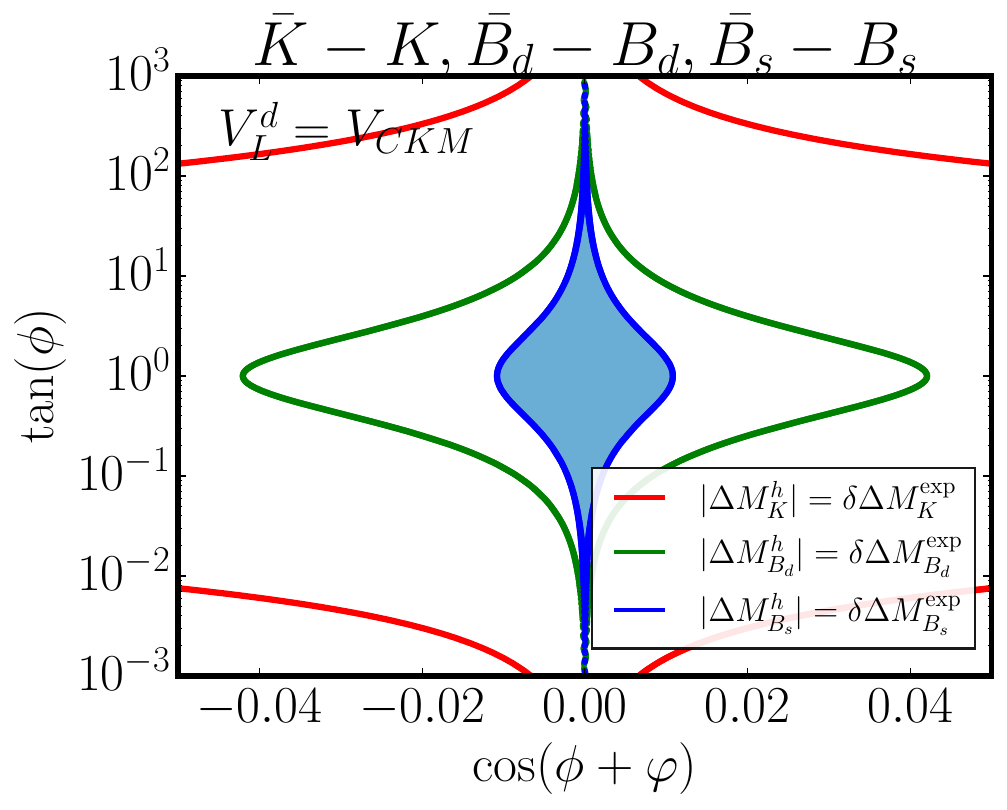}    \includegraphics[width=0.49\linewidth]{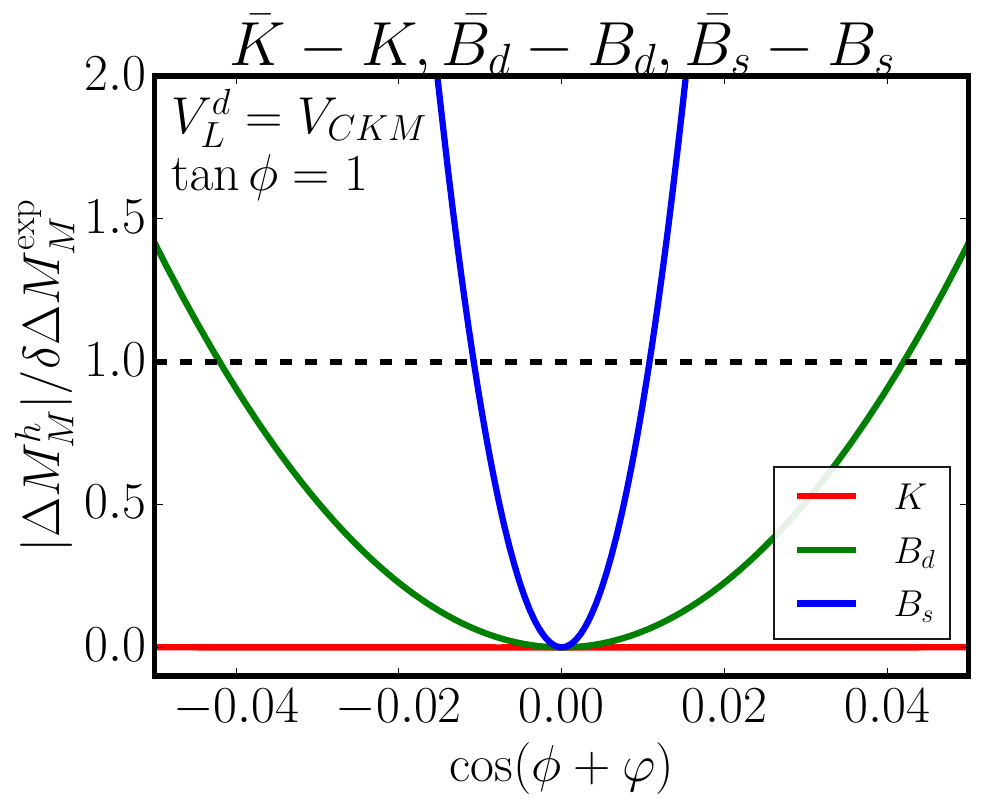}
    \caption{(a) Contours on the $\cos(\phi+\varphi)$--$\tan\phi$ plane fulfilling $100\%$ of the mass differences $\delta\Delta M_M ^\text{exp}$, for mesons $K$ (red), $B_d$ (green) and $B_s$ (blue). As indicated in the figure, outwards of the contours represents excluded region of the parameter space. When combined all meson oscillation constraints, the only remaining allowed region is indicated by the blue contour, Here, the mixing pattern is $V_L^d=V_\text{CKM}$ and $V_L^u=\mathbf{1}$. (b) Mass difference as a function of $\cos(\phi+\varphi)$ for mesons $K$ (red), $B_d$ (green) and $B_s$ (blue), fixing $\tan \phi=1$. The dashed black line denotes $|\Delta M_M^h|=\delta \Delta M_M^\text{exp}$.}
\label{fig:higgs_ii}
\end{figure}

In Fig.~(\ref{fig:higgs_ii}a), we analyze the parameter space dependence of the Higgs-mediated contribution $\Delta M^h_M$ in the $\cos(\phi+\varphi)$--$\tan\phi$ plane. Exclusion regions are derived for three meson systems: $K$ (red curve, $\Delta M_K^h/\delta\Delta M_K^\text{exp} = 1$), $B_d$ (green curve, $\Delta M_{B_d}^h/\delta\Delta M_{B_d}^\text{exp} = 1$), and $B_s$ (blue curve, $\Delta M_{B_s}^h/\delta\Delta M_{B_s}^\text{exp} = 1$). The combined constraints from all three systems yield the final allowed region (blue contour), dominated by the stringent $B_s$ oscillation bounds, which restrict $|\cos(\phi+\varphi)| < 0.01$. In contrast, $K$ and $B_d$ systems permit significantly broader ranges of $\cos(\phi+\varphi)$.

In Fig.~(\ref{fig:higgs_ii}b), we fix $\tan\phi = 1$ to isolate the $\cos(\phi+\varphi)$--$\Delta M^h_M$ parameter space for individual meson systems. Contribution profiles are shown for $K$ (red), $B_d$ (green), and $B_s$ (blue) mesons,  with curves representing their respective relative contribution to meson oscillation experimental bounds.

\textbf{Scenario (iii):} Now, we turn our attention to an intermediate case, with a less trivial factorization of the CKM matrix. This scenario has non-negligible contributions to three neutral meson-antimeson oscillations, $D$, $B_d$ and $B_s$. However, this choice of parametrization suppresses the $K-\bar{K}$ oscillation. We propose the following quark mixing matrices:
\begin{equation}\label{VLUcase3}
    V_L^u =
\left(
\scalemath{0.8}{
\begin{array}{ccc}
 0.01658\, -\,0.9996\,\mathrm{i} & (-0.1531\,-\,0.7210\, \mathrm{i})\times10^{-4} & (-0.3146\,+\,0.1240\, \mathrm{i})\times 10^{-2} \\
 (-0.9453\,-\,0.2054\, \mathrm{i})\times10^{-4} & -0.7306\,-\,0.6812\,\mathrm{i} & -0.02221\,-\,0.0350 \,\mathrm{i} \\
 (-0.3326\,+\,0.06068\, \mathrm{i})\times 10^{-2} & 0.03055\, +\,0.02801\, \mathrm{i} & -0.5421\,-\,0.8389\, \mathrm{i} \\
\end{array}
}
\right),
\end{equation}
    and
\begin{equation}
\label{VLDcase3}
    V_L^d =\left(
\scalemath{0.8}{
\begin{array}{ccc}
 0.01614\, -\,0.9740\,\mathrm{i} & 0.1644\, +\,0.1532\,\mathrm{i} & (0.1928\, +\,0.4572\,\mathrm{i})\times 10^{-3} \\
 0.003768\, -\,0.2248 \,\mathrm{i} & -0.7126\,-\,0.6643\,\mathrm{i}
 & (0.4807\, +\,0.1342\,\mathrm{i})\times 10^{-3} \\
 -0.1590\times10^{-7}\,-\,0.5663 \,\mathrm{i}\,\times10^{-3} & (0.04655\, -\,0.4242\, \mathrm{i})\times 10^{-3} & -0.5426\,-\,0.8397 \,\mathrm{i} \\
\end{array}
}
\right),
\end{equation}
and this configuration produces intermediate constraints over the scalar mixing parameters $\phi$ and $\varphi$.

\begin{figure}[h]
    \centering
    \includegraphics[width=0.49\linewidth]{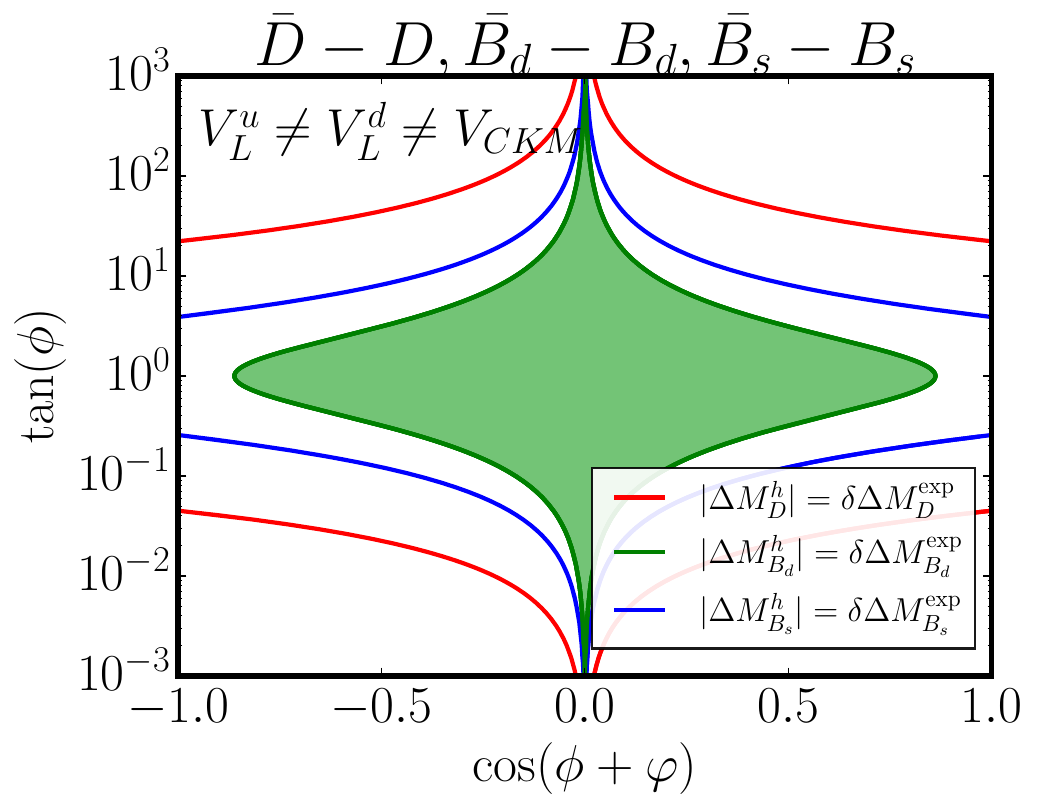}
    \includegraphics[width=0.49\linewidth]{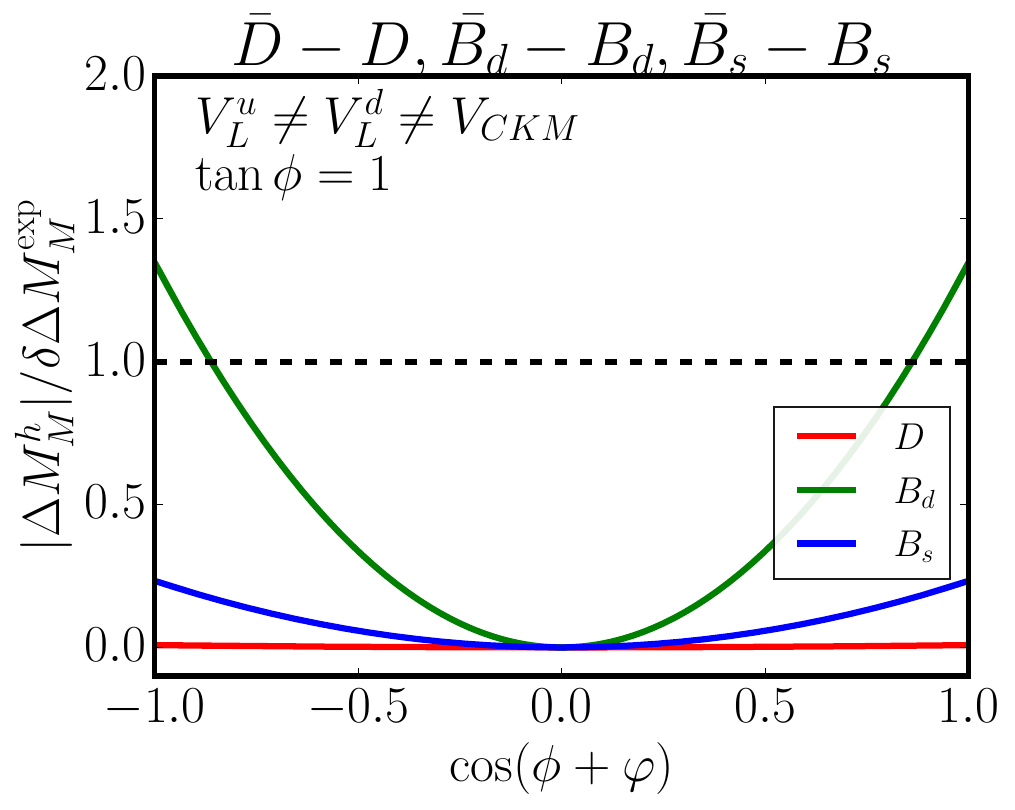}
    \caption{(a) Contours on the $\cos(\phi+\varphi)$--$\tan\phi$ plane fulfilling $100\%$ of the mass differences $\delta\Delta M_M ^\text{exp}$, for mesons $D$ (red), $B_d$ (green) and $B_s$ (blue) (K oscillation allows all presented plane.). As indicated in the figure, outwards of the contours represents excluded region of the parameter space. Here, the mixing pattern  $V_L^u$ and $V_L^d$ are explicitly provided in Eq.~\eqref{VLDcase3}. (b) Mass difference as a function of $\cos(\phi+\varphi)$ for mesons $D$ (red), $B_d$ (green) and $B_s$ (blue), fixing $\tan \phi=1$. The dashed black line denotes $|\Delta M_M^h|=\delta \Delta M_M^\text{exp}$.}
    \label{fig:higgs_iii}
\end{figure}

In Fig.~(\ref{fig:higgs_iii}a), we analyze the parameter space dependence of the Higgs-mediated contribution $\Delta M^h_M$ in the $\cos(\phi+\varphi)$--$\tan\phi$ plane. Exclusion regions are derived for three meson systems: $D$ (red curve, $\Delta M_D^h/\delta\Delta M_D^\text{exp} = 1$), $B_d$ (green curve, $\Delta M_{B_d}^h/\delta\Delta M_{B_d}^\text{exp} = 1$), and $B_s$ (blue curve, $\Delta M_{B_s}^h/\delta\Delta M_{B_s}^\text{exp} = 1$). The combined constraints from all three systems yield the final allowed region (green contour), dominated by the stringent $B_d$ oscillation bounds, which restrict $|\cos(\phi+\varphi)| < 0.9$. In contrast, $D$ and $B_d$ systems permit significantly broader ranges of $\cos(\phi+\varphi)$.

In Fig.~(\ref{fig:higgs_iii}b), we fix $\tan\phi = 1$ to isolate the $\cos(\phi+\varphi)$--$\Delta M^h_M$ parameter space for individual meson systems. Contribution profiles are shown for $D$ (red), $B_d$ (green), and $B_s$ (blue) mesons, with curves representing their respective relative contributions to meson oscillation experimental bounds.  

Now, that we analyzed the alignment limit of the 331 and its dependence on the mixing pattern of the quarks for three different scenarios, we will analyze the same scenarios in light of the new  gauge boson $Z'$ and how each scenario affects the bounds over its mass. 

\section{$Z^{\prime}$ mass estimation using meson oscillation bounds}
\label{sec:FCNC_Zp}

After describing the scalar sector contribution to meson oscillation parameters, it is time for the $Z'$ boson. The  gauge sector of the model  is composed by the standard gauge bosons plus $Z^{\prime}$, two new charged gauge bosons $W^{\prime \pm}$ and two non-hermitian neutral gauge bosons $U^0$ and $U^{0 \dagger}$. For the development of this sector, we refer the reader to Ref.~\cite{Long:1995ctv}. Neglecting mixing among the standard neutral gauge bosons, $Z$ and $Z^{\prime}$, the FCNC processes get intermediated by the neutral scalars and $Z^{\prime}$. The  interactions of $Z^{\prime}$ with the quarks that provide FCNC are given  by

\begin{eqnarray}
{\cal L}^{Z^{\prime}}_u&\supset&\frac{g}{6c_W} \bar{\hat{u}}_{b_L}\left[ \frac{6-6s^2_W}{\sqrt{3-4s^2_W}} (V^u_L)^*_{b3}(V^u_L)_{3a}\right]\gamma^\mu \hat u_{a_L}Z^{\prime},  \\
{\cal L}^{Z^{\prime}}_d&\supset&\frac{g}{6c_W} \bar{\hat{d}}_{b_L}\left[ \frac{6-6s^2_W}{\sqrt{3-4s^2_W}} (V^d_L)^*_{b3}(V^d_L)_{3a}\right]\gamma^\mu \hat d_{a_L}Z^{\prime},
\label{NC}
\end{eqnarray}
where $s_W=\sin \theta_W$ and $c_W=\cos \theta_W$ with $\theta_W$ being the standard Weinberg angle, and $g$ is the weak gauge coupling constant. From this, we write the  effective Lagrangian contributing to meson $M(\bar{q}^bq^a)$ transitions as
\begin{eqnarray}
    \mathcal{L}^{M}_{Z', \text{eff}} &=& \frac{g^2c_W^2}{(3-4s_W^2)M_{Z'}^2} \left| \left(V_L^{u,d}\right)^*_{b3}\left(V_L^{u,d}\right)_{3a} \right|^2 |\bar{q}^b_L \gamma^\mu q^a_L|^2 .
\end{eqnarray}

Consequently, the respective mass differences are given by
\begin{eqnarray}
    \Delta M_{M}^{Z'} &=& \frac{4\sqrt{2}G_F c_W^4}{3-4s_W^2} \left|  \left(V_L^{u,d}\right)^*_{b3}\left(V_L^{u,d}\right)_{3a} \right|^2 \frac{M_Z^2}{M_{Z^{\prime}}^2} f_{M}^2 B_{M} m_{M} \times r_\text{RG},
\end{eqnarray}
where $r_\text{RG}$ accounts for the renormalization group evolution of Wilson coefficients due to the running of the strong coupling constant. At leading order, $r_\text{RG} = 1$, while the one-loop corrections are detailed in Appendix~\ref{app:RGcorr}.

\begin{figure}[h]
    \centering
    \includegraphics[width=0.75\linewidth]{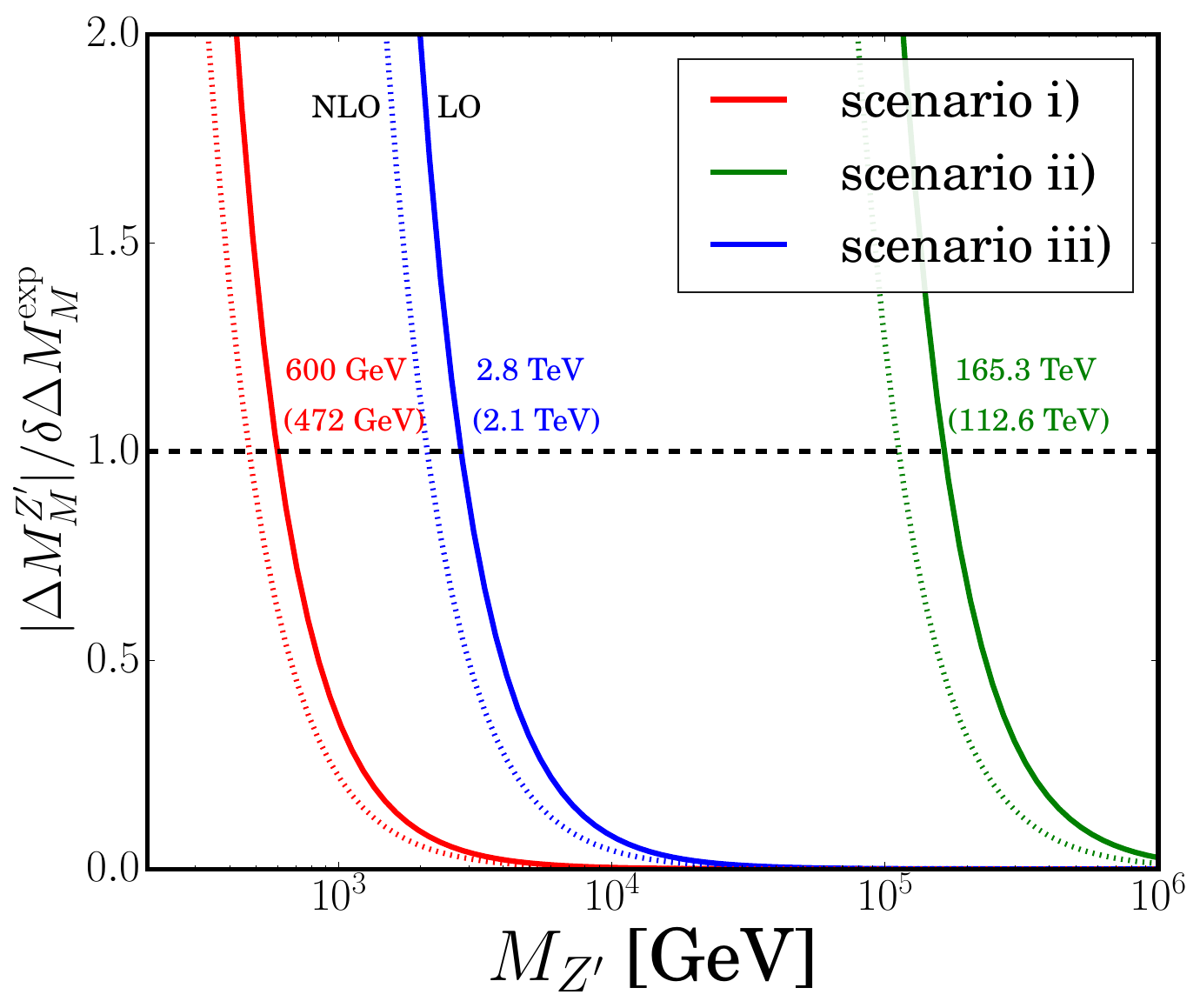}
    \caption{Lower bounds for $M_{Z'}$, presented as the mass difference $\Delta M_M^{Z'}$ as a function of $M_{Z'}$ of the most restrictive meson system of each scenario: $D, B_s$ and $B_d$ for scenarios (i) (red), (ii) (green) and (iii) blue. The full lines represents the LO contributions and the dotted lines represents the NLO contributions.}
    \label{fig:Zprime_bounds}
\end{figure}

After outlining the formalism of the $Z'$-mediated mass splitting $\Delta M_M^{Z'}$, we analyze the three mixing scenarios \textbf{i)}, \textbf{ii)}, and \textbf{iii)} defined in Sec.~\ref{sec:FCNC_higgs}. Our results, displayed in Fig.~\ref{fig:Zprime_bounds}, map the experimental bounds on the $Z'$ mass ($M_{Z'}$) derived from meson oscillation data in Table~\ref{tab:data_table} for leading order (LO) represented as full lines and next-to-leading order (NLO) represented as dotted lines.  

\textbf{Scenario i)} (red curve): The $Z'$ contribution to $D-\bar{D}$ mixing ($\Delta M_D$) yields the weakest constraint. $M_{Z'} \gtrsim 600\;\text{GeV}$ at LO and $M_{Z'} \gtrsim 472\;\text{GeV}$ at NLO, contrasting with conventional expectations \cite{Buras:2012dp,Cogollo:2012ek,Queiroz:2016gif,Okada:2016whh,CarcamoHernandez:2022fvl, deJesus:2023lvn,Oliveira:2022dav,Oliveira:2022dav,Buras:2023ldz}. This suppression arises because $V_\text{CKM}^\dagger = V_L^u$ eliminates FCNC in the down-quark sector, leaving only up-type quark processes. Given the comparatively weak experimental limits on $D-\bar{D}$ oscillations (Table~\ref{tab:data_table}), this scenario permits a remarkably light $Z'$.  

\textbf{Scenario ii)} (green curve): For $B_s-\bar{B}_s$ mixing ($\Delta M_{B_s}$), the $Z'$ mass bound tightens to $M_{Z'} \gtrsim 165.3\;\text{TeV}$ at LO and $112.6$ TeV at NLO, aligning with the literature \cite{Buras:2012dp,Cogollo:2012ek,Queiroz:2016gif,Okada:2016whh,CarcamoHernandez:2022fvl, deJesus:2023lvn,Oliveira:2022dav,Oliveira:2022dav,Buras:2023ldz}. Here, $V_\text{CKM} = V_L^d$ reintroduces FCNCs in the down-quark sector, subjecting $Z'$ couplings to the experimental constraints on $B_s$ oscillations. This explains the dramatic increase in $M_{Z'}$ relative to \textbf{scenario i)}.  

\textbf{Scenario iii)} (blue curve): The hybrid mixing pattern affecting $B_d-\bar{B}_d$ oscillations ($\Delta M_{B_d}$) yields an intermediate bound of $M_{Z'} \gtrsim 2.8\;\text{TeV}$  at LO and $2.1$ TeV at NLO. This scenario introduces FCNCs across both up- and down-type quarks, with $B_d$ observables dominating the constraints.  

Table~\ref{tab:zprimemass} summarizes these results. They highlight how CKM parametrization choices in 331 models critically modulate FCNC visibility, thereby reshaping $Z'$ mass limits.

\begin{table}[t]
    \centering
    \begin{tabular}{|c|c|c|c|}
        \hline
        \diagbox{\textbf{Quark mixing}}{\textbf{Scenario}} & \textbf{(i)} & \textbf{(ii)} & \textbf{(iii)} \\ 
        \hline
        $V_L^u$ & $V_\text{CKM}^\dagger$ & $\mathbf{1}$ & Eq.~\ref{VLUcase3} \\  
        $V_L^d$ & $\mathbf{1}$ & $V_\text{CKM}$ & Eq.~\ref{VLDcase3} \\
        \hline
        \textbf{Meson} & \textbf{(i)} & \textbf{(ii)} & \textbf{(iii)} \\
        \hline
        $D-\bar{D}$ & $M_{Z'} \geq 600.0(473.0)$ GeV& $-$ & $M_{Z'} \geq 582.3(459.4)$ GeV \\  
        $K-\bar{K}$ & $-$ & $M_{Z'} \geq 6.5(4.8)$ TeV & $M_{Z'} \geq 5.0 (4.9)$ GeV \\  
        $B_d-\bar{B_d}$ & $-$ & $M_{Z'} \geq 49.7(34.7)$ TeV & $M_{Z'} \geq 2.8 (2.1)$ TeV  \\  
        $B_s-\bar{B_s}$ & $-$ & $M_{Z'} \geq 165.3(112.6)$ TeV & $M_{Z'} \geq 1.7(1.3)$ TeV \\  
        \hline
    \end{tabular}
    \label{tab:zprimemass}
    \caption{Lower bounds on $M_{Z'}$ for each type of meson oscillation for scenarios \textbf{i)}, \textbf{ii)} and \textbf{iii)}. The limits are present at LO and NLO outside and inside the parenthesis, respectively. }
\end{table}

\section{Summary and conclusion}
\label{sec:summary}

We have conducted a comprehensive analysis of flavor constraints in the 331 model, focusing on the interplay between quark mixing parametrizations, scalar-sector alignment, and $Z'$ mass limits. Our study establishes two important results:

\begin{itemize}
    \item \textbf{Weakest bound over the $Z'$ mass in the literature:} By analyzing three distinct quark mixing scenarios, we demonstrate that $Z'$ mass bounds vary dramatically—from $\sim\!600(473)\;\text{GeV}$ to $\sim\!165\;\text{TeV}$—depending on how $V_L^{u,d}$ rotations distribute FCNC effects across meson systems (Fig.~\ref{fig:Zprime_bounds}, Table II). The extreme suppression in \textbf{scenario i)} ($V_L^u=V_\text{CKM}^\dagger$) arises from confinement of FCNCs to $D$-meson oscillations, whose experimental uncertainties are $\sim\!100\times$ larger than those for $B_s$ systems.    
    \item \textbf{Scalar alignment limit in 331 models:} We identify and characterize a novel alignment condition, $\cos(\phi + \varphi) = 0$, where the SM-like Higgs decouples from FCNCs (Figs.~\ref{fig:higgs_i}--\ref{fig:higgs_iii}). This mechanism differs fundamentally from 2HDM alignment due to the 331 model's extended gauge structure and non-universal quark family assignments. The alignment limit's viability depends critically on the variant of the model and on the $V_L^{u,d}$ parametrization, with $B_s$-meson data imposing $|\cos(\phi+\varphi)| < 0.01$ in \textbf{scenario ii)}.
\end{itemize}

These findings challenge the prevailing assumption that 331 models universally require multi-TeV $Z'$ masses. By exploiting quark rotation ambiguities—particularly configurations suppressing FCNCs in the down-quark sector—we reduce the $Z'$ mass floor to $\sim\!600(473)\;\text{GeV}$, well within LHC reach. Crucially, this flexibility arises naturally from the model's gauge structure. 

We  emphasize that our constraint on  $M_{Z^{\prime}}$ stem from a highly aggressive approach, wherein new physics was assumed to lie within the experimental error margins. These lower bounds can be further relaxed by adopting an alternative approach, where new physics is required to fall within the range of theoretical uncertainties.  

On the other side, this work provides the first combined analysis of SM-like Higgs and $Z'$-mediated FCNCs, revealing their synergistic constraints. Future studies should explore correlations with collider signatures (e.g., $Z' \to \mu^+\mu^-$ resonances) and electroweak precision observables, which may further constrain or corroborate the viability of light $Z'$ scenarios, while in the Higgs sector exploring the new alignment limit and exploring the constraints over $\cos(\phi+\varphi)$ over the actual experimental limits of signal strength.

\section*{Acknowledgments}
P. E. is supported by CNPq grant No. 151612/2024-2. 
C. A. S. P.  is supported by the CNPq research grants No. 311936/2021-0. J. P. P. is supported by grant  PID2022-\allowbreak 126224NB-\allowbreak C21 and  "Unit of Excellence Maria de Maeztu 2020-2023'' award to the ICC-UB CEX2019-000918-M  funded by MCIN/AEI/\allowbreak 10.13039/\allowbreak 501100011033, 
also supported by the European Union's through the
Horizon 2020 research and innovation program (Marie
Sk{\l}odowska-Curie grant agreement 860881-HIDDeN) and the Horizon
Europe research and innovation programme (Marie Sk{\l}odowska-Curie
Staff Exchange grant agreement 101086085-ASYMMETRY). 
It also receives support  from grant 2021-SGR-249 (Generalitat de Catalunya).
.

\appendix
\section{Scalar potential}
\label{app:scalarpotential}
The most general potential that preserves lepton number and is invariant under $Z_2$ involves the following set of terms\footnote{ For previous works developing the scalar sector of the model, see: \cite{Pal:1994ba,Long:1997vbr,Tully:1998wa,Ponce:2002sg,Palcu:2013sfa,Diaz:2003dk} },

\begin{eqnarray} 
V(\eta,\rho,\chi)&=&\mu_\chi^2 \chi^2 +\mu_\eta^2\eta^2
+\mu_\rho^2\rho^2+\lambda_1\chi^4 +\lambda_2\eta^4
+\lambda_3\rho^4+ \nonumber \\
&&\lambda_4(\chi^{\dagger}\chi)(\eta^{\dagger}\eta)
+\lambda_5(\chi^{\dagger}\chi)(\rho^{\dagger}\rho)+\lambda_6
(\eta^{\dagger}\eta)(\rho^{\dagger}\rho)+ \nonumber\\
&&\lambda_7(\chi^{\dagger}\eta)(\eta^{\dagger}\chi)
+\lambda_8(\chi^{\dagger}\rho)(\rho^{\dagger}\chi)+\lambda_9
(\eta^{\dagger}\rho)(\rho^{\dagger}\eta) \nonumber \\
&&- \dfrac{f}{\sqrt{2}} \epsilon^{ijk} \eta_i \rho_j \chi_k +  \mathrm{H.c.} \, .
\label{PPLN}
\end{eqnarray}

The Higgs spectrum of this potential is very complex and is  investigated in several papers. Here we use the notation and  results  of Ref. \cite{Pinheiro:2022bcs}. As we are going to consider  the contributions of the neutral scalars to flavor changing processes, for pedagogical reasons we revise the neutral scalar sector of the model following, exactly, the analysis in Ref. \cite{Pinheiro:2022bcs} where only the neutral scalars $\eta^0$, $\rho^0$ and $\chi^{\prime 0}$ develop vev different from zero and, as usual, we shift them in the following way ,
\begin{equation}
  \eta^0\,\,, \rho^0\,\,, \chi^{\prime 0} =\frac{1}{\sqrt{2}}(v_{\eta\,, \rho\,,\chi^{\prime}}+R_{_{\eta\,, \rho\,,\chi^{\prime}}}+iI_{_{\eta\,, \rho\,,\chi^{\prime}}})\,.
\end{equation}
The set of  minimal equations that guarantee the potential develop a  minimum is given by
\begin{eqnarray}
 &&\mu^2_\chi +\lambda_1 v^2_{\chi^{\prime}} +
\frac{\lambda_4}{2}v^2_\eta  +
\frac{\lambda_5}{2}v^2_\rho-\frac{f}{2}\frac{v_\eta v_\rho}
{ v_{\chi^{\prime}}}=0,\nonumber \\
&&\mu^2_\eta +\lambda_2 v^2_\eta +
\frac{\lambda_4}{2} v^2_{\chi^{\prime}}
 +\frac{\lambda_6}{2}v^2_\rho -\frac{f}{2}\frac{v_{\chi^{\prime}} v_\rho}
{ v_\eta} =0,
\nonumber \\
&&
\mu^2_\rho +\lambda_3 v^2_\rho + \frac{\lambda_5}{2}
v^2_{\chi^{\prime}} +\frac{\lambda_6}{2}
v^2_\eta-\frac{f}{2}\frac{v_\eta v_{\chi^{\prime}}}{v_\rho} =0.
\label{mincond} 
\end{eqnarray}

Concerning the CP-odd scalars and adopting the basis $(I_{\chi^{\prime}}, I_\eta, I_\rho)$, we obtain the following mass matrix for these scalars:
\begin{equation}
M_I^2=
\begin{pmatrix}
fv_\eta v_\rho/4v_{\chi^{\prime}} &  f v_\rho/4 &  f v_\eta/4\\
f v_\rho/4 &  fv_{\chi^{\prime}} v_\rho/4v_\eta & f v_{\chi^{\prime}}/4\\
 f v_\eta/4 &  f v_{\chi^{\prime}}/4 &  fv_{\chi^{\prime}} v_\eta/4v_\rho
\end{pmatrix}.
\label{MI}
\end{equation}
The diagonalization of this mass matrix must provides two massless pseudoscalars that will be the Goldsotne eaten by the standard gauge boson,$Z$, and the new one, $Z^{\prime}$. The remaining pseudoscalar, $A$, is a combination of $I_{\chi^{\prime}}$,  $I_\eta$ and $I_\rho$. In the approximation $v_{\chi^{\prime}}\gg v_\eta\,,\,v_\rho$, we have, 
\begin{equation}
    A \approx  \frac{v_\rho}{\sqrt{v^2_\eta + v^2_\rho}} I_\eta +\frac{v_\eta}{\sqrt{v^2_\eta + v^2_\rho}}I_\rho ,
\end{equation}
with  $m^2 _A=\frac{f v_{\chi^{\prime}}}{4}(\frac{v_\eta v_\rho}{v^2_{\chi^{\prime}}}+\frac{v_\eta}{v_\rho}+\frac{v_\rho}{v_\eta})$.

Concerning the CP-even scalars, in the basis $(R_{\chi^{\prime}}, R_\eta,R_\rho)$, we have the following mass matrix,
\begin{equation}
M_R^2=
\begin{pmatrix}
\lambda_1 v^2_{\chi^{\prime}}+fv_\eta v_\rho/4v_{\chi^{\prime}} & \lambda_4 v_{\chi^{\prime}}  v_\eta/2- f v_\rho/4 &  \lambda_5 v_{\chi^{\prime}}  v_\rho/2- f v_\eta/4\\
\lambda_4 v_{\chi^{\prime}}  v_\eta/2- f v_\rho/4 & \lambda_2 v^2_\eta+ fv_{\chi^{\prime}} v_\rho/4v_\eta & \lambda_6 v_\eta v_\rho/2-f v_{\chi^{\prime}}/4\\
\lambda_5 v_{\chi^{\prime}}  v_\rho/2- f v_\eta/4 &  \lambda_6 v_\eta v_\rho/2-f v_{\chi^{\prime}}/4 &  \lambda_3 v^2_\rho+ fv_{\chi^{\prime}} v_\eta/4v_\rho
\end{pmatrix}.
\label{masseven}
\end{equation}
It is impossible to diagonalize analytically this mass matrix. We, then, work in the decoupling limit that is when $R_\eta$ and $R_\rho$ decouple from $R_{\chi^{\prime}}$. This is achieved by imposing the elements $1 \times 2$ and $1 \times 3$ to be null. This happens for $\lambda_5=\lambda_4 \frac{v^2_\eta}{v^2_\rho}$. In this case we have 
\begin{equation}
M_R^2=
\begin{pmatrix}
\lambda_1 v^2_{\chi^{\prime}}+fv_\eta v_\rho/4v_{\chi^{\prime}} & 0 &  0\\
0 & \lambda_2 v^2_\eta+ fv_{\chi^{\prime}} v_\rho/4v_\eta & \lambda_6 v_\eta v_\rho/2-f v_{\chi^{\prime}}/4\\
0 &  \lambda_6 v_\eta v_\rho/2-f v_{\chi^{\prime}}/4 &  \lambda_3 v^2_\rho+ fv_{\chi^{\prime}} v_\eta/4v_\rho
\end{pmatrix}.
\label{massevenII}
\end{equation}
Diagonalizing this matrix by means of a rotation in the space of $(R_\eta\,,\,R_\rho$), we may display the results by means of the mixture:
\begin{eqnarray}
&&H^{\prime}= R_\chi,\nonumber \\
  && h= \cos \varphi R_\eta +\sin \varphi R_\rho ,\nonumber \\
  &&
  H=-\sin \varphi R_\eta +\cos \varphi R_\rho,
  \label{S1}
\end{eqnarray}
with $H$ and $H ^{\prime}$ having their  masses proportional to $v_{\chi^{\prime}}$. The scalar  $h$ must recover the properties of the standard Higgs, as discused in Ref. \cite{Pinheiro:2022bcs}. In the decoupling limit, $H^{\prime}$ does not  contribute to the FCNC processes involving the standard quarks. Thus, in what matter to FCNC intermediated by neutral scalars, we just  need to worry with $h$, $H$ and $A$.  

\section{Coefficients of effective lagrangians}
\label{app:coefficients}
Besides the interaction (\ref{eq:Ycaseh1}), concerning $h$, the Yukawa interactions among $H$ and $A$ and the physical
quarks which are given by
\begin{eqnarray}
  {\cal L}^{H}_Y&=& 2  \frac{\sin (\phi -\varphi)}{|\sin 2\phi|}(V^d_L)_{3a}^* (V^d_L)_{b3} \frac{(m_{\text{up}})_a}{v} \bar{\hat{u}}_{b_L} \hat u_{a_R}H\\
 &+&  2    \frac{\sin (\phi +\varphi)}{|\sin 2\phi|} (V^d_L)_{3a}^* (V^d_L)_{b3} \frac{(m_{\text{down}})_a}{v} \bar{\hat{d}}_{b_L} \hat d_{a_R}H  + \text{H.c.}\,, 
\end{eqnarray}
and
\begin{eqnarray}
  {\cal L}_Y^{A}&=&i\left(\tan{\phi} +  \cot{\phi}     \right)(V^u_L)_{3a}^* (V^u_L)_{b3} \frac{(m_{\text{up}})_a}{v}  \bar{\hat{u}}_{b_L}\hat u_{a_R} A \nonumber \\
&+& i\left(\tan{\phi}  -\cot{\phi} \right)(V^d_L)_{3a}^* (V^d_L)_{b3}\frac{(m_{\text{down}})_a}{v}  \bar{ \hat{d}}_{b_L}\hat d_{a_R} A  + \text{H.c.}\,.
\end{eqnarray}
 Consequently, a meson $M$ composed of a bound state of a quark-antiquark pair $\bar{\hat{q}}_a \hat{q}_b$ oscillates with its antiparticle through the following effective interactions:
\begin{equation}
    \mathcal{L}_{h, \text{eff}}^{M} =\left(\frac{2\cos(\varphi+\phi)}{m_h\sin(2\phi)}\right)^2\left(\bar{\hat{q}}_a\left(C_{M,h}^R P_R - C_{M,h}^L P_L\right)\hat{q}_b\right)^2,
\end{equation}
\begin{equation}
    \mathcal{L}_{H, \text{eff}}^{M} =\left(\frac{2}{m_H\sin(2\phi)}\right)^2\left(\bar{\hat{q}}_a\left(C_{M,H}^R P_R - C_{M,H}^L P_L\right)\hat{q}_b\right)^2,
\end{equation}
and
\begin{equation}
    \mathcal{L}_{A, \text{eff}}^{M} =-\frac{1}{m_A^2}\left(\bar{\hat{q}}_a\left(C_{M,A}^R P_R - C_{M,A}^L P_L\right)\hat{q}_b\right)^2.
\end{equation}
The meson mass differences induced by $h$ are given in Eq. \ref{contribution_h}. The contributions of $H$ and $A$ are given by
\begin{align}
\Delta  m_{M}^{H}= \frac{8}{\sin^2(2\phi)}\frac{m_{M} B_{M} f_{M}^2}{m_{H}^2}\left[\frac{5}{24} \operatorname{Re}\left[\left(C_{M,H}^L\right)^2+\left(C_{M,H}^R\right)^2\right]\left(\frac{m_{M}}{m_{q_a}+m_{q_b}}\right)^2 \right. \nonumber \\  + \left( 2 \operatorname{Re}\left[C_{M,H}^L C_{M,H}^R\right] \left(\frac{1}{24} +\frac{1}{4}\left(\frac{m_{M}}{m_{q_a}+m_{q_b}}\right)^2\right)\right],
\end{align}
and
\begin{align}
\Delta m_{M}^{A}=-\frac{2 m_{M} B_{M} f_{M}^2}{m_{A}^2}\left[\frac{5}{24} \operatorname{Re}\left[\left(C_{M,A}^L\right)^2+\left(C_{M,A}^R\right)^2\right]\left(\frac{m_{M}}{m_{q_a}+m_{q_b}}\right)^2 \right. \nonumber \\ + \left( 2 \operatorname{Re}\left[C_{M,A}^L C_{M,A}^R\right]\left(\frac{1}{24}+\frac{1}{4}\left(\frac{m_{M}}{m_{q_a}+m_{q_b}}\right)^2\right)\right].
\end{align}
The coefficients $C_{M,h}^{R,L},\,\,C_{M,H}^{R,L}$ and $C_{M,A}^{R,L}$ are given by the following expressions:
\begin{eqnarray}
    C_{D,h}^R &=& \frac{m_c}{v}  \left(V_L^u\right)_{32}^* \left(V_L^u\right)_{13}, \nonumber \\
    C_{D,h}^L &=& \frac{m_u}{v}   \left(V_L^u\right)^*_{13} \left(V_L^u\right)_{32}, \nonumber \\
    C_{K,h}^R &=& \frac{m_d}{v} \left(V_L^d\right)^*_{31} \left(V_L^d\right)_{23}, \nonumber \\
    C_{K,h}^L &=&  \frac{m_s}{v}  \left(V_L^d\right)^*_{23} \left(V_L^d\right)_{31}, \nonumber \\
    C_{B_d,h}^R &=& \frac{m_d}{v} \left(V_L^d\right)_{31}^* \left(V_L^d\right)_{33}, \nonumber \\
    C_{B_d,h}^L &=&  \frac{m_b}{v}  \left(V_L^d\right)^*_{33} \left(V_L^d\right)_{31}, \nonumber \\
    C_{B_s,h}^R &=& \frac{m_s}{v} \left(V_L^d\right)_{32}^* \left(V_L^d\right)_{33}, \nonumber \\
    C_{B_s,h}^L &=&  \frac{m_b}{v}  \left(V_L^d\right)^*_{33} \left(V_L^d\right)_{32},
\end{eqnarray}
\begin{eqnarray}
    C_{D,H}^R &=& \frac{m_c}{v} \sin (\phi-\varphi)\left(V_L^u\right)_{32}^* \left(V_L^u\right)_{13}, \nonumber \\
    C|_{D,H}^L &=& \frac{m_u}{v} \sin (\phi-\varphi)\left(V_L^u\right)^*_{13} \left(V_L^u\right)_{32}, \nonumber \\
    C_{K,H}^R &=& \frac{m_d}{v} \sin (\phi+\varphi) \left(V_L^d\right)^*_{31} \left(V_L^d\right)_{23}, \nonumber \\
    C_{K,H}^L &=&  \frac{m_s}{v} \sin (\phi+\varphi) \left(V_L^d\right)^*_{23} \left(V_L^d\right)_{31}, \nonumber \\
    C_{B_d,H}^R &=& \frac{m_d}{v} \sin (\phi+\varphi) \left(V_L^d\right)_{31}^* \left(V_L^d\right)_{33}, \nonumber \\
    C_{B_d,H}^L &=&  \frac{m_b}{v} \sin (\phi+\varphi) \left(V_L^d\right)^*_{33} \left(V_L^d\right)_{31}, \nonumber \\
    C_{B_s,H}^R &=& \frac{m_s}{v} \sin (\phi+\varphi) \left(V_L^d\right)_{32}^* \left(V_L^d\right)_{33}, \nonumber \\
    C_{B_s,H}^L &=&  \frac{m_b}{v} \sin (\phi+\varphi)  \left(V_L^d\right)^*_{33} \left(V_L^d\right)_{32},
\end{eqnarray}
\begin{eqnarray}
    C_{D,A}^R &=& \frac{m_c}{v} (\tan\phi+\cot\phi) \left(V_L^u\right)_{32}^* \left(V_L^u\right)_{13}, \nonumber \\
    C_{D,A}^L &=& \frac{m_u}{v} (\tan\phi+\cot\phi)  \left(V_L^u\right)^*_{13} \left(V_L^u\right)_{32}, \nonumber\\
    C_{K,A}^R &=& \frac{m_d}{v} (\tan\phi-\cot\phi) \left(V_L^d\right)^*_{31} \left(V_L^d\right)_{23}, \nonumber \\
    C_{K,A}^L &=&  \frac{m_s}{v} (\tan\phi-\cot\phi) \left(V_L^d\right)^*_{23} \left(V_L^d\right)_{31}, \nonumber \\
    C_{B_d,A}^R &=& \frac{m_d}{v} (\tan\phi-\cot\phi) \left(V_L^d\right)_{31}^* \left(V_L^d\right)_{33}, \nonumber \\
    C_{B_d,A}^L &=&  \frac{m_b}{v} (\tan\phi-\cot\phi) \left(V_L^d\right)^*_{33} \left(V_L^d\right)_{31}, \nonumber \\
    C_{B_s,A}^R &=& \frac{m_s}{v}(\tan\phi-\cot\phi) \left(V_L^d\right)_{32}^* \left(V_L^d\right)_{33}, \nonumber  \\
    C_{B_s,A}^L &=&  \frac{m_b}{v} (\tan\phi-\cot\phi) \left(V_L^d\right)^*_{33} \left(V_L^d\right)_{32}.
\end{eqnarray}

\section{Renormalization group evolution of the Wilson coefficient at one-loop}
\label{app:RGcorr}

The renormalization group (RG) evolution of the $\Delta F = 2$ Wilson coefficients from the $Z'$ mass scale to the relevant meson scales.

From equation (21), the effective Lagrangian after integrating out the $Z'$ is:
\begin{equation}
\mathcal{L}_{\text{eff}}^{\Delta F=2} = \frac{g'^2}{8M_{Z'}^2} \sum_{i,j} (V_L^{d})_{3i}^* (V_L^{d})_{3j} (\bar{d}_i \gamma^\mu P_L d_j)^2
\end{equation}

This gives the Wilson coefficient at the matching scale $\mu = M_{Z'}$:
\begin{equation}
C_1(M_{Z'}) = \frac{g'^2}{8M_{Z'}^2} (V_L^{d})_{3i}^* (V_L^{d})_{3j}
\end{equation}

The RG evolution of the Wilson coefficient is governed by:
\begin{equation}
\mu \frac{d}{d\mu} C_1(\mu) = \gamma_1(\alpha_s) C_1(\mu)
\end{equation}

The solution at leading-log order is:
\begin{equation}
C_1(\mu_L) = \left[\frac{\alpha_s(\mu_H)}{\alpha_s(\mu_L)}\right]^{\gamma_1^{(0)}/(2\beta_0)} C_1(\mu_H)
\end{equation}
where $\mu_H = M_{Z'}$ and $\mu_L$ is the low-energy scale.

For the operator $\mathcal{O}_1 = (\bar{d}_i \gamma^\mu P_L d_j)^2$:
\begin{equation}
\gamma_1^{(0)} = 6 C_F = 6 \times \frac{4}{3} = 8
\end{equation}

The QCD beta function coefficient:
\begin{equation}
\beta_0 = \frac{11N_c - 2n_f}{3} = \frac{33 - 2n_f}{3}
\end{equation}

For $M_{Z'} > m_t$, we must include threshold matching (supposing there are only the standard quarks):
\begin{itemize}
\item For $\mu > m_t$: $n_f = 6$, $\beta_0 = 7$
\item For $m_b < \mu < m_t$: $n_f = 5$, $\beta_0 = 23/3$
\item For $m_c < \mu < m_b$: $n_f = 4$, $\beta_0 = 25/3$
\item For $\mu < m_c$: $n_f = 3$, $\beta_0 = 9$
\end{itemize}

The complete RG factor for $B_{d,s}$ mixing ($\mu_L \approx m_b$) is:

For $M_{Z'} \gg m_t$:
\begin{equation}
r_{RG} = \left[\frac{\alpha_s(m_t)}{\alpha_s(m_b)}\right]^{12/23} \left[\frac{\alpha_s(M_{Z'})}{\alpha_s(m_t)}\right]^{4/7}
\end{equation}

The one-loop RG solution for $\alpha_s$ is:
\begin{equation}
\alpha_s(\mu_2) = \frac{\alpha_s(\mu_1)}{1 + \frac{\beta_0 \alpha_s(\mu_1)}{2\pi} \ln\left(\frac{\mu_2}{\mu_1}\right)}
\end{equation}

Using $\alpha_s(M_Z) = 0.118$, $\alpha_s(m_b) = 0.22$, $\alpha_s(m_t) = 0.108$:
\begin{align}
M_{Z'} = 1 \text{ TeV}: & \quad \sqrt{r_{RG}} \approx 0.80 \\
M_{Z'} = 10 \text{ TeV}: & \quad \sqrt{r_{RG}} \approx 0.73 \\
M_{Z'} = 100 \text{ TeV}: & \quad \sqrt{r_{RG}} \approx 0.69
\end{align}

Including RG effects, equation (22) becomes:
\begin{equation}
\Delta M_{12}^\text{new}= r_{RG} \times \Delta M_{12}^\text{old} 
\end{equation}

Then, the bounds for NLO are:
\begin{equation}
M_{Z^{\prime}}=\sqrt{\frac{ r_{RG} \times \frac{4\sqrt{2}G_F c_W^4}{3-4s_W^2} \left|  \left(V_L^{u,d}\right)^*_{b3}\left(V_L^{u,d}\right)_{3a} \right|^2 M_Z^2 f_{M}^2 B_{M} m_{M}}{ \Delta M_{12}^\text{exp}}}
\end{equation}

This relaxes the bounds on $M_{Z'}$ by:
\begin{equation}
M_{Z'}^{\text{with RG}} =\sqrt{r_{RG}} M_{Z'}^{\text{without RG}}
\end{equation}

Then, 
\begin{align}
M_{Z'} = 1 \text{ TeV} &\to  0.8 \text{ TeV} \\
M_{Z'} = 10 \text{ TeV} &\to  7.3 \text{ TeV} \\
M_{Z'} = 100 \text{ TeV} &\to  68.8 \text{ TeV}
\end{align}

\bibliography{references}
\end{document}